\documentclass[apj]{emulateapj}
\usepackage{apjfonts,epsf}
\newcommand{\Msun}{M_{\sun}}

\begin{document}

\slugcomment{Submitted to ApJ \today}
\shortauthors{GNEDIN ET AL.}
\shorttitle{HALO CONTRACTION EFFECT}

\title{Halo Contraction Effect in Hydrodynamic Simulations of Galaxy Formation}

\author{ Oleg Y. Gnedin\altaffilmark{1},
         Daniel Ceverino\altaffilmark{2},
         Nickolay Y. Gnedin\altaffilmark{3,4},
         Anatoly A. Klypin\altaffilmark{5},\\
         Andrey V. Kravtsov\altaffilmark{4},
         Robyn Levine\altaffilmark{6},
         Daisuke Nagai\altaffilmark{7},
         Gustavo Yepes\altaffilmark{8}
}

\altaffiltext{1}{Department of Astronomy, University of Michigan,
   Ann Arbor, MI 48109; \mbox{\tt ognedin@umich.edu}}
\altaffiltext{2}{Racah Institute of Physics, The Hebrew University, 
   Jerusalem 91904, Israel}
\altaffiltext{3}{Particle Astrophysics Center, Fermi National Accelerator Laboratory, 
   Batavia, IL 60510}
\altaffiltext{4}{Department of Astronomy and Astrophysics,
   Kavli Institute for Cosmological Physics,
   The University of Chicago, Chicago, IL 60637}
\altaffiltext{5}{Department of Astronomy, New Mexico State University, 
   Las Cruces, NM 88003}
\altaffiltext{6}{CITA, Toronto, ON, M5S 3H8, Canada}
\altaffiltext{7}{Department of Physics, Yale University, New Haven, CT 06520}
\altaffiltext{8}{Grupo de Astrof\'isica, Universidad Aut\'onoma de Madrid, 
   Madrid E-28049, Spain}

\begin{abstract}
The condensation of gas and stars in the inner regions of dark matter
halos leads to a more concentrated dark matter distribution.  While
this effect is based on simple gravitational physics, the question of
its validity in hierarchical galaxy formation has led to an active
debate in the literature.  We use a collection of several
state-of-the-art cosmological hydrodynamic simulations to study the
halo contraction effect in systems ranging from dwarf galaxies to
clusters of galaxies, at high and low redshift.  The simulations are
run by different groups with different codes and include hierarchical
merging, gas cooling, star formation, and stellar feedback.  We show
that in all our cases the inner dark matter density increases relative
to the matching simulation without baryon dissipation, at least by a
factor of several.  The strength of the contraction effect varies from
system to system and cannot be reduced to a simple prescription.  We
present a revised analytical model that describes the contracted mass
profile to an rms accuracy of about 10\%.  The model can be used to
effectively bracket the response of the dark matter halo to baryon
dissipation.  The halo contraction effect is real and must be included
in modeling of the mass distribution of galaxies and galaxy clusters.
\end{abstract}


\section{Adiabatic Contraction in the Context of Hierarchical Galaxy Formation}
    
Dissipationless cosmological simulations predict that virialized halos
of dark matter are described by an approximately universal density
profile \citep{dubinski_carlberg91, navarro_etal97, navarro_etal10,
  moore_etal98}.  Although non-baryonic dark matter exceeds normal
baryonic matter by a factor of $\Omega_{\rm dm}/\Omega_{\rm b} \approx
5$ on average in the universe, the gravitational field in the central
regions of galaxies can be dominated by stars and gas.  In the
hierarchical galaxy formation picture, cosmic gas dissipates its
thermal energy, condenses towards the halo center, and forms stars.
In the process, dark matter particles are pulled inward and increase
their central density.

The response of dark matter to baryonic infall has traditionally been
calculated using the model of adiabatic contraction
\citep{eggen_etal62, zeldovich_etal80, barnes_white84}.  The present
form of the standard adiabatic contraction (SAC) model was introduced
by \citet{blumenthal_etal86} and \citet{ryden_gunn87}.  This model
assumes that a spherically symmetric halo can be thought of as a
sequence of concentric shells, made of particles on circular orbits,
which homologously contract while conserving the angular momentum.
With these assumptions, the final location $r_f$ of the shell
enclosing mass $M_{\rm dm}(r)$, which was at radius $r$ before the
contraction, can be calculated by knowing only the initial ($M_{{\rm
    b},i}$) and final ($M_{{\rm b},f}$) baryon mass profile:
\begin{equation}
  \left[ M_{\rm dm}(r) + M_{{\rm b},i}(r) \right] r =
  \left[ M_{\rm dm}(r) + M_{{\rm b},f}(r_f) \right] r_f.
  \label{eq:sac}
\end{equation}
Since the baryon mass typically increases at the inner radii as a
result of dissipation ($M_{{\rm b},f}(r) > M_{{\rm b},i}(r)$), the
final radius of the shell is smaller than the initial radius, $r_f <
r$.  That is, the halo contracts.

The effect of contraction of the dark matter distribution is important
for many galactic studies: for modeling the mass profiles of galaxies
and clusters of galaxies, for studying star formation feedback on the
galactic structure, for abundance matching of galaxies and dark matter
halos, for predicting possible signatures of dark matter annihilation,
and others.  This importance has led to a lively debate in the recent
literature, both theoretical and observational, on the validity of the
effect (we discuss it in more detail in Section~\ref{sec:literature}).
Hierarchical galaxy formation is considerably more complex than the
simple picture of quiescent cooling in a static spherical halo.  Every
halo is assembled via a series of mergers of smaller units, with the
cooling of gas and contraction of dark matter occurring separately in
each progenitor.  Some objects may undergo dissipationless merging
after the gas is exhausted or the cooling time becomes too long.  Any
of these effects may invalidate the assumptions of the SAC model and
possibly its prediction of halo contraction.

The only way to validate halo contraction is by investigating
cosmological hydrodynamic simulations that self-consistently model as
many of these complex processes as possible: gas dissipation, star
formation, and its feedback.  In \citet{gnedin_etal04}, we tested the
SAC model using a suite of such simulations of the formation of one
Milky Way-sized galaxy and eight clusters of galaxies, performed with
the Adaptive Refinement Tree (ART) code.  The comparison of the
matching pairs of simulations with and without gas dissipation showed
that the halos always contracted.  However, the effect was weaker
than that predicted by the SAC model.  We developed a modified adiabatic
contraction (MAC) model based on the modified invariant, $M(\bar{r}) r$,
where $\bar{r}$ is the orbit-averaged radius for particles currently
located at radius $r$:
\begin{equation}
  \left[ M_{\rm dm}(\bar{r}) + M_{{\rm b},i}(\bar{r}) \right] r =
  \left[ M_{\rm dm}(\bar{r}) + M_{{\rm b},f}(\bar{r_f}) \right] r_f.
  \label{eq:mac}
\end{equation}
Although this combination is not strictly conserved during galaxy
formation, it is a simple correction that most accurately predicts the
dark matter profiles in the simulations.  Using the mass within the
average radius approximately accounts for eccentricity of particle
orbits in a cosmologically-assembled halo.  Averaging over particle
orbits in radial bins, we found a mean relation between $\bar{r}$ and
$r$ in the range $10^{-3}\lesssim r/r_{\rm vir}\lesssim 1$:
\begin{equation}
  {\bar{r} \over r_{\rm vir}} = A \left({r \over r_{\rm vir}}\right)^w,
  \label{eq:rrave}
\end{equation}
where $r_{\rm vir}$ is the virial radius.  This power-law dependence
reflects typical energy and eccentricity distributions of particles in
cold dark matter halos in the dissipationless simulations.  The
parameters $A$ and $w$ varied from halo to halo and from epoch to
epoch.  For the systems at hand we found the mean values $A \approx
0.85$ and $w \approx 0.8$, which we used as fiducial parameters of the
model.  We showed that the MAC model predicts the halo mass profile to
10\%-20\% accuracy.

Since 2004, several numerical and observational studies have evaluated
the accuracy of the MAC model.  We discuss these studies in
Section~\ref{sec:literature}.  In particular,
\citet{gustafsson_etal06} found different amounts of contraction in
their four simulated halos and suggested that different combinations
of the model parameters may best fit individual profiles.  Following the
idea that the MAC model can be directly tested, in this paper we
assemble a large collection of hydrodynamic simulations, of different
systems run with different codes, to study the systematics of the
contraction effect.  We consider the effect of baryons only on the
spherically-averaged radial halo distribution, although the halo shape
is also affected \citep[e.g.,][]{kazantzidis_etal04, abadi_etal10,
zemp_etal11}.

\section{Debate in the Literature}
  \label{sec:literature}

\subsection{Theoretical Studies}
  
Early controlled simulations confirmed the halo contraction effect
\citep{sellwood_mcgaugh05, choi_etal06, colin_etal06}.  

Recently, \citet{duffy_etal10} considered 67 galactic and galaxy group
halos in three cosmological simulations with different feedback
prescription with the Smooth Particle Hydrodynamics (SPH) code Gadget.
They found the contraction effect in all cases, but each with a
different amount of contraction.  Similarly to
\citet{gustafsson_etal06}, they explored the best-fit distribution of
parameters $(A,w)$ and found them to be in the range $0.2 - 0.4$ at
$z=0$.  These parameter values may have been systematically lowered by
the details of the fitting.

\citet{abadi_etal10} studied 13 galaxy-sized halos simulated with the
SPH code GASOLINE.  The simulations included radiative cooling of
cosmic gas above $10^4$~K but no cooling below $10^4$~K and no star
formation.  They found that in each case the halo contracted but by a
smaller amount than predicted by either SAC or MAC model.  However,
the weaker contraction may be an artifact of the assumed gas physics,
because their Figure~8 shows that the earlier simulations by the same
authors \citep{abadi_etal03, meza_etal03}, which included star
formation and feedback, agree much better with the MAC model.

\citet{pedrosa_etal09, pedrosa_etal10} and \citet{tissera_etal10}
studied 6 halos in high-resolution SPH simulations with Gadget, with
different prescriptions for star formation and feedback.  They found
halo contraction in all cases but again weaker than predicted by
either SAC or MAC model.  They suggest that the dynamical formation
history may affect the amount of contraction in each individual halo.

In contrast to these studies, in a simulation of 3 galaxies with
Gadget, \citet{johansson_etal09} found that the dark matter mass
contained with a fixed radius of 2 kpc increased at high redshift,
reached a peak at $z \approx 3$, and then declined continuously to
$z=0$.  The central density decreased by up to a factor of two between
$z=3$ and $z=0$, reaching the same value as in a matching
dissipationless simulation and thus effectively canceling the
contraction.  \citet{johansson_etal09} attribute the reduction of the
dark matter density to gravitational heating by infalling dense
stellar clumps.  In their run most of the baryons in the halo were
converted into stars at high redshift (stellar mass is 13\% of the
virial mass), creating massive stellar satellites, which may be too
effective at pushing out dark matter.  The evolution of the central
density may also be affected by a spurious numerical effect due to
two-body scattering of massive particles.  This scattering would alter
the profile on the two-body relaxation time, which at 2 kpc is about 9
Gyr in their highest-resolution run, and is shorter in other runs.

The SPH simulation of a large galaxy by \citet{romanodiaz_etal08} with
constrained, rather than cosmological, initial conditions yielded an
even more unexpected result.  While they found the steepening of the
dark matter profile at high redshift, the profile subsequently
flattened to a constant-density core within the inner 3 kpc by $z=0$,
reversing a contraction into an expansion.  Such flattening is likely
to be a numerical relaxation effect.  Based on the information in
their paper, we estimate that there are about 9000 particles within 3
kpc, which have a two-body relaxation time of about 14 Gyr.  Thus the
interaction of these particles can alter the true profile within the
age of the universe.

Two new GASOLINE simulations, one for a dwarf galaxy
\citep{governato_etal10} and one for a Milky Way-sized galaxy
\citep{guedes_etal11}, reach different conclusions on halo
contraction.  In the first case the central dark matter density is
reduced, while in the second it is enhanced.  Both simulations use the
same strong blastwave feedback, which suppresses gas cooling around
the star-forming region for several Myr and creates powerful outflows
that remove a significant fraction of baryons from the halo (70\% in
the first run and 30\% in the second run).  The difference can be
traced to a higher density threshold for star formation in the
\citet{governato_etal10} simulation, which means the supernova energy
release is more concentrated and creates rapid potential fluctuations
near the center.  As \citet{pontzen_governato11} emphasize, the
decrease of the inner dark matter density is achieved by repeated
fluctuations following bursts of star formation, because a single
outflow event cannot cause a significant change
\citep[e.g.,][]{gnedin_zhao02}.  However, note that the
\citet{governato_etal10} result depends directly on the adopted
parameters (their lower-threshold version in
\citealt{governato_etal07} showed halo contraction instead of
expansion) and thus it is one of many possible outcomes.

Overall, cosmological simulations performed by different authors with
very different codes and physics input agree that the contraction
effect is present, but at a weaker level than suggested by the SAC
model and with a significant variation from system to system.

\subsection{Observational Studies}

Observational studies have generally provided evidence for halo
contraction.  \citet{schulz_etal10} studied the mass distribution of a
stacked sample of 75,000 elliptical galaxies from the Sloan Digital
Sky Survey, combining weak lensing analysis in the outer regions of
the halo with measurements of the stellar velocity dispersion in the
inner regions.  They found that halo contraction is required to
explain the large and significant mass excess over the dissipationless
NFW profile, in all luminosity bins.  The MAC model gave excellent
predictions to the data, while the SAC model overestimated the excess.
An alternative explanation of the observed mass excess, by changing
the stellar mass-to-light ratio, would require doubling the stellar
mass relative to the Kroupa IMF.

\citet{auger_etal10} studied the mass distribution of 53 early-type
galaxies from the Sloan Lens ACS Survey, using a combination of
stellar velocity dispersion measurements, strong lensing, and weak
lensing.  They found that the data favor a heavy stellar IMF and that
halo contraction predicted by the MAC model improves the fit relative
to the SAC model or to the case without halo contraction.

\citet{minor_kaplinghat08} investigated the effect of halo contraction
on the statistics of strong gravitational lensing.  They find that the
contraction effect enhances the total lensing probability (similarly
found by \citealt{rozo_etal08}) and increases the ratio of double
images relative to quad images and lensing cusps.  In particular,
without contraction naked cusp configurations would become dominant at
angular separation as small as 2.5\arcsec, which is inconsistent with
the data.

While early X-ray studies of the mass distribution in galaxy clusters
\cite[e.g.,][]{zappacosta_etal06} disfavored halo contraction,
\citet{democles_etal10} studied the mass distribution in fossil galaxy
groups (otherwise known as ``X-Ray Overluminous Elliptical Galaxies'',
see \citealt{vikhlinin_etal99}) and found that halo contraction
slightly improves the model fit.

\citet{seigar_etal08} studied the mass distribution in the M31 galaxy
and found that halo contraction is required to fit the inner regions.
Without contraction, they cannot find any set of halo parameters that
would reproduce the peak of the optical rotation curve at 10 kpc from
the center, even with a very large halo mass.  In fact, the best fit
requires the SAC model, while the fiducial MAC model would need an
unreasonably large concentration parameter of the dark matter halo.

\citet{napolitano_etal11} derived the mass profile of a giant
elliptical galaxy M84 using radial velocity measurements of planetary
nebulae and field stars.  The inclusion of the halo contraction
effect, as parametrized by the MAC model, has allowed to derive the
best-fit parameters of the dark halo to be in excellent agreement with
the predictions of cosmological simulations.  It is the first time
that such agreement was achieved with the planetary nebulae kinematics.

\citet{dutton_etal07, dutton_etal11} constructed semi-analytical
models that simultaneously fit the luminosity function of galaxies and
the Tully-Fisher/Faber-Jackson scaling relations.  They found a
preference for halo contraction for early-type galaxies, but a need
for the opposite (halo expansion) for late-type galaxies.  This
conclusion is subject to additional uncertainty in the stellar
mass-to-light ratio or the stellar IMF.

\citet{benson_bower10} investigated extensive observational data
(galaxy sizes, clustering, luminosity function, mass-metallicity
relation, Tully-Fisher relation, cosmic star formation history) using
an improved version of the semi-analytical code {\sc galform} and a
flexible prescription for halo contraction using the MAC model.  Their
best-fit model required strong contraction ($A=0.74$, $w=0.92$; see
also their Figure 7).  Overall, they found a range of MAC model
parameters ($A=0.74-0.96$, $w=0.81-0.99$) matching the multiple data
constraints.

\section{What controls the amount of contraction?}

\citet{gnedin_etal04} suggested that the eccentricity of dark matter
particle orbits may be responsible for the weaker contraction effect
seen in the simulations, relative to the SAC model.  During the
process of baryon condensation particle orbits can further deform,
although \citet{debattista_etal08} find that this effect is almost
completely reversible if the baryons are subsequently removed (for
example, by strong stellar feedback).  In addition, during violent
changes of the gravitational potential, such as those during galaxy
mergers or interactions, the orbital structure may differ from that in
a dynamical equilibrium.  Then the orbital action variables are not
strictly conserved and a first-principles calculation of the
contraction effect becomes even more difficult.

Our original power-law approximation for the average orbit radius
$\bar{r}$ was derived as an average of several dissipationless
simulations of galaxy-sized and cluster-sized halos.  Hydrodynamic SPH
simulations of \citet{gustafsson_etal06} confirmed that
Equation~(\ref{eq:rrave}) is an excellent fit in the inner parts of
halos.  They fit each of their four simulated halos with an
independent set of the parameters $A$ and $w$, and found them to lie
in the range $A = 0.74-0.83$, $w = 0.69-0.81$.  The values of the
parameters in our MAC model are consistent with these ranges.
However, the results of \citet{gustafsson_etal06} indicated that each
simulated system may have somewhat differing amounts of contraction.

\begin{figure}[t]
\vspace{-0.4cm}
\centerline{\epsfxsize=3.6in \epsffile{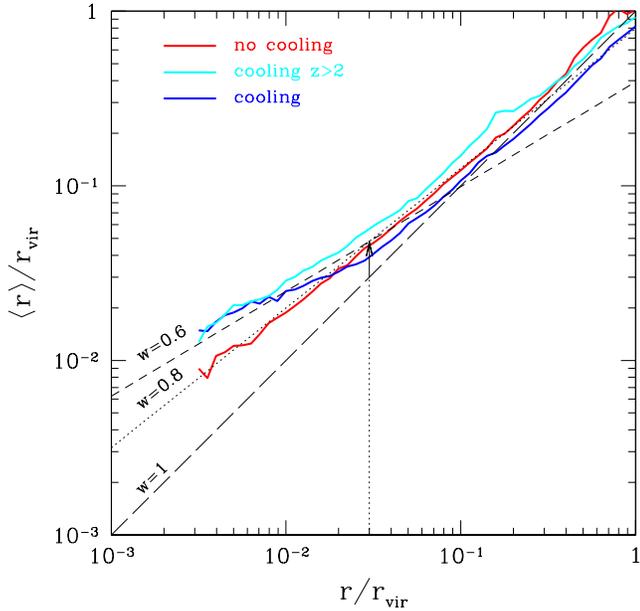}}
\vspace{-0.4cm}
\caption{Average orbital radius vs. instantaneous radius of dark
  matter particles in a dissipationless and two dissipational runs of
  a cluster-sized halo at $z=0$.  This halo is described in the second
  line in Table~\ref{tab:sim} and in Section~\ref{sec:clu}.  The size
  of the smallest resolution element is $\Delta = 1.7\times 10^{-3}\,
  r_{\rm vir}$.  In one simulation the gas cooling is
  suppressed at $z<2$.  Dotted and dashed lines show the slopes in the
  inner halo, with $w=0.8$ and $w=0.6$, respectively.  Long-dashed
  line shows the relation in the SAC model.  Vertical dotted line
  marks new pivot radius $r_0$ in Equation~\ref{eq:rrave0}.  Vertical
  arrow illustrates an adjustment by factor of 1.6 required for new
  normalization $A_0$.}
\vspace{0.2cm}
  \label{fig:rave}
\end{figure}

To better understand the orbital dependence, we investigated the
$\bar{r}-r$ relation in our hydrodynamic run for ``Cluster 6'' (shown
in Figures 1 and 2 in \citealt{gnedin_etal04}).  We computed a proxy
for the average orbital radius, $\bar{r} \approx \langle r \rangle$,
by integrating orbits in a static NFW potential normalized to the
maximum circular velocity $V_{\rm max}$ measured in the simulation.
Note that while the normalization of the profile correctly takes into
account the baryon condensation, the shape of the potential may
deviate from the NFW form in the dissipative runs.
Figure~\ref{fig:rave} shows that while the dissipationless run is well
characterized by our fiducial slope $w \approx 0.8$, the run with
cooling and star formation shows a more complex relation, with
departures from a single power law.  The outer regions, between $0.1
r_{\rm vir}$ and $r_{\rm vir}$, have a steep slope $w \approx 1$ (the
virial radius is calculated at the overdensity of 180 with respect to
the mean cosmic density).  The inner regions have much flatter slope,
$w \approx 0.5$, similar to the values reported by
\citet{gustafsson_etal06}.  In these inner regions, $r < 0.03 r_{\rm
vir}$, the contribution of baryons to the mass profile becomes
significant and the contraction effect should be most important.  Thus
for the purpose of calibrating the contraction model, we can restrict
the fit of a power-law slope just to the inner regions.

Interestingly, in the run in which cooling and star formation were
turned off at $z<2$, the $\bar{r}-r$ relation in the outer halo is
closer to that in the dissipationless run than in the full dissipative
run.  In the inner halo, the slope is again flatter, $w \approx 0.6$.
In this run the stellar fraction within the virial radius is half of
that in the full run and is closer to the observed fraction in galaxy
clusters.  After the cooling was stopped, the gravitational potential
in the inner halo probably did not evolve and allowed the particle
orbits to come to dynamical equilibrium.  Therefore, the contraction
effect may depend not only on the final amount of baryon condensation
but also on the duration of this process.

Since the $\bar{r}-r$ relation evolves during the process of
dissipation, a MAC model with fixed parameters $A$ and $w$ cannot
accurately describe the contraction effect.  Instead, following
\citet{gustafsson_etal06}, we can search for the best-fitting values
of the model parameters for each simulated halo and then analyze their
distribution.

Figure~\ref{fig:rave} also shows that if we are to vary $w$, the
normalization $A$ would have to vary correspondingly because
Equation~(\ref{eq:rrave}) is anchored at $r_{\rm vir}$, far from the
region of interest in the inner halo.  We can reduce this correlated
variation of $A$ by shifting the pivot of the $\bar{r}-r$ relation to
some inner radius, $r_0$.  The correlation between $w$ and $A$, which
we derive for the simulations described below, is minimized for $r_0
\approx 0.03 \, r_{\rm vir}$.  Therefore, we redefine the relation of
the MAC model as
\begin{equation}
  {\bar{r} \over r_0} = A_0 \left({r \over r_0}\right)^w.
  \label{eq:rrave0}
\end{equation}
We use the subscript ``0'' to differentiate this normalization
parameter $A_0$ from its original version in \citet{gnedin_etal04}
defined by Equation~(\ref{eq:rrave}).  We fix the pivot at $r_0 = 0.03
\, r_{\rm vir}$ in all discussion that follows.  Only the
normalization $A_0$ is affected by this shift of the pivot point from
$r_{\rm vir}$ to $r_0$, the slope $w$ remains invariant.  The SAC
model is still characterized by $A_0=w=1$.

\subsection{The role of model parameters}

To understand the effect of varying the parameters $A$ and $w$ on the
amount of contraction, consider the inner regions of a halo.  The
contraction factor $y \equiv r_f/r$ obeys Equation~(\ref{eq:mac}),
which after dividing both sides by $M_i(\bar{r}) r_f$ becomes
\begin{equation}
  {1 \over y} = 1-f_b + {M_b(\overline{ry}) \over M_i(\bar{r})},
  \label{eq:y}
\end{equation}
where $f_b$ is the baryon fraction within the virial radius of the
halo of interest.  The average cosmic baryon fraction is $f_b \approx
0.17$, but a given halo may have a different baryon fraction.  Here
$M_i(r) \equiv M_{{\rm dm},i}(r) + M_{{\rm b},i}(r)$ is the total
initial mass, the sum of dark matter and baryons.  Typically, before
contraction the dark matter profile shadows the total mass
distribution, $M_{{\rm dm},i}(r) \approx (1-f_b) \, M_i(r)$.

In the inner halo described by an NFW profile with the scale radius
$r_s$, the initial enclosed mass is $M_i(r) \propto r^2$ at $r \ll
r_s$.  In this region, we can define the average logarithmic slope of
the final baryon density profile: $\rho_b \propto r^{-\nu}$, which
lies in the range $1 < \nu < 3$ ($\nu \approx 2$ is typical;
\citealt{koopmans_etal09}).  Then the final baryon mass profile
$M_b(r) \propto r^{3-\nu}$, and Equation (\ref{eq:y}) becomes
\begin{equation}
  {1 \over y} = 1-f_b + {M_b(\bar{r}) \over M_i(\bar{r})}\; y^{w(3-\nu)}.
  \label{eq:yi}
\end{equation}

The enhancement factor of the dark matter mass profile is given by
Equation~(A11) of \citet{gnedin_etal04} and is most easily evaluated
at the contracted radius $r_f = r y(r)$:
\begin{equation}
  F_M(ry) \equiv {M_{{\rm dm},f}(ry) \over M_{{\rm dm},i}(ry)} 
          = {M_{{\rm dm},i}(r) \over M_{{\rm dm},i}(ry)} \approx {1 \over y(r)^2}.
  \label{eq:fm}
\end{equation}
The last approximation is valid only in the inner region
where $M_i(r) \propto r^2$.  For arbitrary $\nu$ and $w$,
Equation~(\ref{eq:yi}) is transcendental and must be solved
numerically.  However, a good second-order approximation 
is described in Appendix.

To make concrete calculations using Equations~(\ref{eq:yi}) and
(\ref{eq:fm}), we must specify the ratio of the final baryon enclosed
mass to the initial total enclosed mass.  Let $r_e$ be the radius
where the final baryon mass equals the initial dark matter mass:
$M_b(r_e) \equiv M_{{\rm dm},i}(r_e)$.  For the Milky Way galaxy, this
radius is about 10 kpc, and thus a good choice is $r_e \approx 0.05\,
r_{\rm vir}$.  It then follows that $M_b(\bar{r})/M_i(\bar{r}) =
(1-f_b) \, (\bar{r}/r_e)^{1-\nu}$.

\begin{figure}[t]
\vspace{-0.4cm}
\centerline{\epsfxsize=3.6in \epsffile{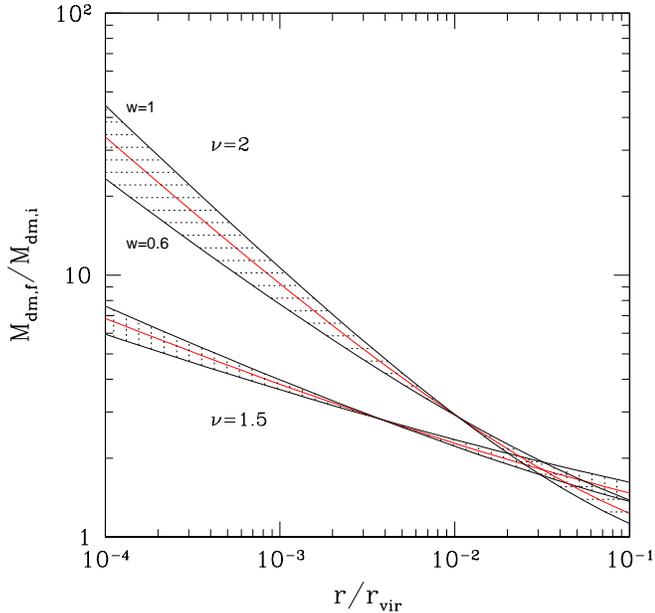}}
\vspace{-0.4cm}
\caption{Enhancement of the enclosed dark matter mass given by
  Equation~(\ref{eq:fm}) for the inner halo.  The top shaded region
  corresponds to the baryon density slope $\nu=2$, the bottom region
  to $\nu=1.5$.  Each region is bounded by the lines of the
  contraction models with the MAC parameter $w=1$ (upper) and $w=0.6$
  (lower).  The middle line is for $w=0.8$.  The parameter $A_0$ is set
  to 1.6 in all cases.  The baryon mass profile is normalized to be
  equal to the initial dark matter mass at $r_e = 0.05 \, r_{\rm vir}$.
}
  \label{fig:enhance}
\vspace{0.2cm}
\end{figure}

Figure~\ref{fig:enhance} shows the mass enhancement factor using the
numerical solutions of Equation~(\ref{eq:yi}) for two representative
values of $\nu=2$ and $\nu=1.5$.  In the inner halo, the stronger
baryon dissipation ($\nu=2$ vs. $\nu=1.5$) leads to the stronger
contraction of dark matter.  Larger value of $w$ and smaller value of
$A_0$ indicate stronger contraction effect at small radii.  However,
there is a cross-over point for the lines of different $w$, and the
trend is reversed at the intermediate radii ($0.01-0.1\, r_{\rm
vir}$).  While both $A_0$ and $w$ determine the normalization of $F_M$,
the parameter $w$ determines also the increase of the amount of
contraction with radius.

\begin{table}[t]
\begin{center}
\caption{\sc Simulations and MAC model parameters}
\label{tab:sim}
\begin{tabular}{llllclcc}
\tableline\tableline\\
\multicolumn{1}{l}{System} &
\multicolumn{1}{c}{$M_{\rm vir}$ ($\Msun$)} &
\multicolumn{1}{c}{$A_0$} &
\multicolumn{1}{c}{$w$} &
\multicolumn{1}{c}{rms} &
\multicolumn{1}{c}{$w^a$} &
\multicolumn{1}{c}{rms$^a$} &
\multicolumn{1}{c}{Ref}
\\[2mm] \tableline\\
Cluster      & $3.4\times 10^{14}$ & 1.83 & 1.33 & 0.037 & 1.29 & 0.060 & G04 \\
Cluster      & $3.7\times 10^{14}$ & 1.74 & 0.51 & 0.028 & 0.50 & 0.045 & G04 \\
Cluster      & $3.8\times 10^{14}$ & 1.54 & 0.74 & 0.015 & 0.76 & 0.022 & G04 \\
Cluster      & $2.1\times 10^{14}$ & 2.30 & 1.48 & 0.051 & 1.57 & 0.140 & G04 \\
Cluster      & $1.4\times 10^{14}$ & 1.78 & 0.71 & 0.009 & 0.79 & 0.053 & G04 \\
Cluster      & $2.0\times 10^{14}$ & 1.28 & 0.55 & 0.027 & 0.37 & 0.100 & G04 \\
Cluster      & $1.9\times 10^{14}$ & 1.54 & 1.03 & 0.007 & 1.02 & 0.016 & G04 \\
Cluster      & $1.2\times 10^{14}$ & 1.24 & 0.87 & 0.013 & 0.61 & 0.232 & G04 \\
Group        & $8.6\times 10^{13}$ & 1.42 & 0.67 & 0.009 & 0.55 & 0.032 & N06 \\
Group        & $4.0\times 10^{13}$ & 1.30 & 1.06 & 0.005 & 0.81 & 0.030 & N06 \\
Group        & $7.1\times 10^{13}$ & 1.47 & 0.91 & 0.011 & 0.82 & 0.024 & N06 \\
Group        & $1.1\times 10^{14}$ & 1.43 & 1.11 & 0.023 & 0.98 & 0.036 & N06 \\
Group        & $5.3\times 10^{13}$ & 1.14 & 1.22 & 0.035 & 0.84 & 0.067 & N06 \\
Group        & $7.3\times 10^{13}$ & 2.06 & 0.91 & 0.012 & 1.24 & 0.078 & N06 \\
Group        & $9.8\times 10^{13}$ & 1.59 & 0.85 & 0.017 & 0.84 & 0.057 & N06 \\
Group        & $7.6\times 10^{13}$ & 1.41 & 0.69 & 0.009 & 0.55 & 0.031 & N06 \\
Group        & $6.1\times 10^{13}$ & 0.75 & 1.28 & 0.008 & 0.59 & 0.135 & N06 \\
Group        & $2.1\times 10^{13}$ & 1.12 & 1.29 & 0.020 & 0.90 & 0.048 & N06 \\
Group        & $1.9\times 10^{13}$ & 1.87 & 0.62 & 0.010 & 0.77 & 0.017 & N06 \\
Group        & $1.2\times 10^{14}$ & 1.61 & 0.86 & 0.027 & 0.86 & 0.027 & N06 \\
Galaxy       & $5.5\times 10^{11}$ & 2.65 & 1.38 & 0.009 & 0.85 & 0.107 & G10 \\
Galaxy       & $7.5\times 10^{11}$ & 1.79 & 1.20 & 0.014 & 1.07 & 0.034 & G10 \\
Galaxy       & $2.8\times 10^{11}$ & 1.21 & 0.67 & 0.016 & 0.91 & 0.068 & G10 \\
Galaxy $z=1$ & $8.4\times 10^{11}$ & 2.07 & 0.64 & 0.021 & 0.99 & 0.081 & C09 \\
Galaxy $z=1$ & $4.8\times 10^{10}$ & 2.92 & 0.85 & 0.023 & 1.31 & 0.266 & C09 \\
Galaxy $z=3$ & $3.4\times 10^{11}$ & 1.32 & 1.26 & 0.103 & 1.26 & 0.144 & L08 \\
\tableline
\end{tabular}
\end{center}
\vspace{-0.2cm}
{\bf Note:} $^a$ for fixed $A_0=1.6$.\smallskip\\
{\bf References:} 
  G04 = \citet{gnedin_etal04}, N06 = \citet{nagai06}, 
  G10 = \citet{gottlober_etal10}, C09 = \citet{ceverino_klypin09}, 
  L08 = \citet{levine_etal08}.
\vspace{0.3cm}
\end{table}

\section{Cosmological hydrodynamic simulations}
  \label{sec:sim}

In this section we describe the simulations of galaxy formation that
we use to test the MAC model.  We tried to collect as diverse samples
of simulations and codes as possible and to analyze them at $z=0$, so
as to evaluate the applicability of the model to observations.  We
also consider a few interesting cases at $z=1$ and $z=3$.

We fit both the original set of parameters $A$ and $w$ from Equation
(\ref{eq:rrave}) and the modified set $A_0$ and $w$ from Equation
(\ref{eq:rrave0}).  The latter set for each system is
summarized in Table~\ref{tab:sim}.  

To derive the best-fit parameters we minimize the difference between
the enclosed mass in a radial bin in the dissipative simulation,
$M_{{\rm sim}}$, and the corresponding mass predicted by the MAC
model, $M_{{\rm mod}}$.  For convenience, we define the following
dimensionless difference in radial bin $i$:
\begin{equation}
  z_i \equiv {M_{{\rm mod},i} \over M_{{\rm sim},i}} -1.
\end{equation}
The ``correct'' mass profile in the dissipative simulation is not
known infinitely precisely, and therefore the model should match it
only within certain fidelity, $\sigma_i$.  The best-fit values of the
parameters are obtained by minimizing the $\chi^2$ function
\begin{equation}
  \chi^2 = \sum_i {z_i^2 \over \sigma_i^2}.
\end{equation}

We include two sources of uncertainty in $\sigma_i$.  The first is a
simple Poisson counting error, which we take to be the square root of
the number of particles in the bin, $N_i^{-1/2}$.  The other is
systematic uncertainty in the spherical mass profile, which can arise
from a number of effects such as a triaxiality of the halo, 
an ongoing merger event, or a small mismatch in the output timing
between the dissipational and dissipationless simulations.  In such
cases the deviation in some radial bins between the predicted model
profile from the simulated profile may be large, even if the bin
contains thousands of particles.  We model this uncertainty as a
constant error in each bin, $\sigma_{\mathrm{sys}}$, that puts an
upper limit on the value of $\chi^2$.  This allows us to obtain
controlled, even if only relative, estimates of the confidence
intervals of the model parameters.  The value of
$\sigma_{\mathrm{sys}}$ is set such that for the SAC model ($A_0=w=1$)
it would result in a specified value of $\chi^2$ per degree of freedom
(equal to $N_{\mathrm{sys}}$):
\begin{equation}
  \sigma_{\mathrm{sys}}^2 \equiv {1 \over N_{\mathrm{sys}}} 
                                 {1 \over N_b-2} \sum_i z_i^2,
\end{equation}
where $N_b$ is the number of bins, and the number of degrees of
freedom is $N_b-2$ when we are fitting two model parameters.  Once we
set $\sigma_{\mathrm{sys}}$, it is then kept fixed as we search for the
best-fitting parameters $A$ and $w$.

We take a large enough value, $N_{\mathrm{sys}} = 100$, such
that it would not alter fitting in the bins with fewer than 100
particles, while avoiding large deviations in the more populous bins.
We have verified that choosing any value in the range $10 <
N_{\mathrm{sys}} < 1000$ results in similar model parameters.  The total error
in bin $i$ is
\begin{equation}
  \sigma_i^2 = {1 \over N_i} + \sigma_{\mathrm{sys}}^2.
\end{equation}

As an alternative to minimizing the $\chi^2$ function, we have also
tested a robust estimator $\sum |z_i| / \sigma_i$, which is not
sensitive to distant outliers.  With both methods we obtained
essentially the same best-fit parameters for our simulations.

In order to make the most accurate fit in the region where the
contraction effect is important, we restrict the radial bins included
in the fitting to $r < 0.1\, r_{\rm vir}$.  We allow the parameters to
vary in the range $0 < A_0 \le 3$, $0 < w \le 2$.  Even larger values
would yield models degenerate with those in the chosen range, as it
will become clear later in Figure~\ref{fig:aw}.

\begin{figure}[t]
\centerline{\epsfxsize=3.6in \epsffile{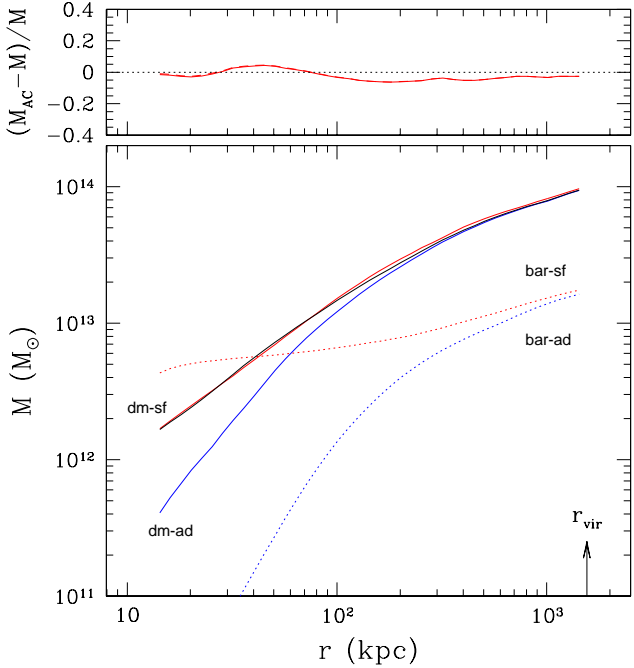}}
\vspace{-0.2cm}
\caption{Contraction of the dark matter profile in a simulated group
  of galaxies at $z=0$, from \citet{nagai06}.  Solid lines show the
  enclosed dark matter mass profile, in the non-radiative (ad) run
  and star formation (sf) run.  Dotted lines show the
  corresponding baryon mass profiles.  Black solid line is the best
  fit of the MAC model, with parameters $A_0=1.61$, $w=0.86$.  Top
  panel shows the mass residuals for the MAC model with freely
  adjustable $A_0$ ({\it solid}) and the MAC model with fixed
  $A_0=1.6$ ({\it dashed}).  In this plot the two lines almost coincide.}
  \label{fig:daisuke}
\vspace{0.2cm}
\end{figure}

\subsection{Clusters and groups of galaxies}
  \label{sec:clu}

The first sample consists of the simulations of 8 galaxy cluster halos
described in \citet{gnedin_etal04} and additional 12 galaxy group
halos by \citet{nagai06} with the same setup.  These are
high-resolution cosmological simulations in the $\Lambda$CDM model
($\Omega_{\rm m}=0.3$, $\Omega_{\Lambda}=0.7$, $\Omega_{\rm b}=0.043$,
$h=0.7$, $\sigma_8=0.9$) performed with the ART code
\citep*{kravtsov99, kravtsov_etal02}.  The simulations have a peak
spatial resolution $\Delta = 3.5$~kpc and dark matter particle mass of
$3.9\times 10^8\, \Msun$.  The virial mass of the systems ranges from
$2\times 10^{13}\, \Msun$ to $4\times 10^{14}\, \Msun$.  Star
formation is implemented using the standard Kennicutt's law and is
allowed to proceed in regions with temperature $T<10^4$~K and gas
density $n_g > 0.1\ \rm cm^{-3}$.  We truncate the inner profiles at
$4 \Delta$ to ensure that the gravitational dynamics is calculated
correctly in the studied region.

Figure~\ref{fig:daisuke} shows the mass profiles for one of the
groups.  The dark matter mass is significantly enhanced in the star
formation run relative to the non-radiative run, by a factor of 4 at the
innermost resolved radius.  The baryons strongly dominate the total
mass at that point.  The MAC model provides an excellent fit to the
contracted dark matter profile, with the parameters ($A_0=1.61$,
$w=0.86$) close to the fiducial values.  The maximum deviation of the
mass profile predicted by the MAC model is 6\%, and the rms
deviation over all bins at radii $r < 0.1\, r_{\rm vir}$ is 3\%.  We
similarly analyzed the other eleven groups and present their best-fit
parameters in the discussion of Figure~\ref{fig:aw}.

\subsection{Individual Galaxies}

\begin{figure}[t]
\centerline{\epsfxsize=3.6in \epsffile{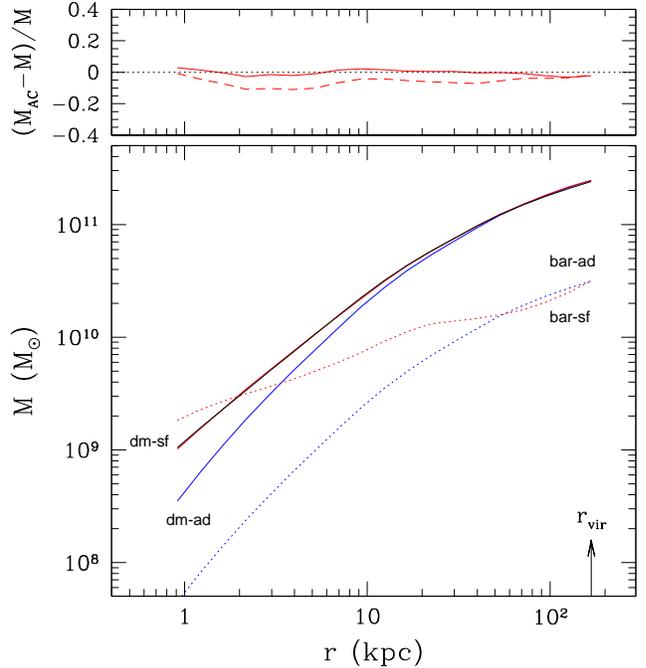}}
\vspace{-0.2cm}
\caption{Contraction of the dark matter profile in a simulated galaxy
  at $z=0$, from \citet{gottlober_etal10}.  Line notation is the same
  as in Figure~\ref{fig:daisuke}.  Black solid line is the best-fit
  MAC model with $A_0=1.79$, $w=1.2$.  Dashed line in top panel shows
  the MAC model prediction with fixed $A_0=1.6$, and best-fitting
  $w=1.07$.}
  \label{fig:gustavo}
\vspace{0.2cm}
\end{figure}

We consider the simulation of three Milky Way-sized galaxies by the
CLUES project
\citep[http://www.clues-project.org;][]{gottlober_etal10,
knebe_etal10}.  The simulation is run using the SPH code Gadget-2.
This code includes standard radiative cooling, star formation, and
supernova feedback.  The force softening length $\epsilon = 0.14$~kpc.
The halos were selected from a large box and resimulated with the
effective mass resolution of $4096^3$ dark matter particles.  In the
highest-resolution halos the particle mass is $3.5\times 10^5\,
\Msun$.  The virial masses of the three halos at $z=0$ are
$(3-8)\times 10^{11}\ \Msun$.  The inner truncation radius is set by
the condition that the local two-body relaxation time exceeds the age
of the universe.

Figure~\ref{fig:gustavo} shows the profile of the most massive of the
three galaxies.  The dark matter mass is enhanced by an order of
magnitude at the innermost radius.  The MAC model with parameters
($A_0=1.79$, $w=1.2$) predicts the dark matter profile to better than
4\% accuracy in any bin, with the rms deviation of only 1.4\%.

We consider also the simulations of a Milky Way-sized galaxy and a
dwarf galaxy at $z=1$ by \citet{ceverino_klypin09}.  These simulations
are run with the ART code with a very different prescription for
stellar feedback than in \citet{nagai06}.  The large galaxy mass is
$8\times 10^{11}\, \Msun$, the dwarf galaxy mass is $5\times 10^{10}\,
\Msun$, both at $z=1$.  The dark matter particle mass is $7.5\times
10^5\, \Msun$ and the peak spatial resolution is 100 comoving pc for
the larger galaxy.  For the smaller galaxy, the dark matter particle
mass is $9.4\times 10^4\, \Msun$ and the peak resolution is 50
comoving pc.
Compared to the non-radiative runs, the dark matter mass is enhanced
by a factor of 8 for the larger galaxy and by a factor of 5 for the
smaller galaxy, at the innermost radius.  The MAC model (with
parameters $A_0=2.07$, $w=0.64$ and $A_0=2.92$, $w=0.85$,
respectively) predicts the dark matter profile to better than 9\%
accuracy, with the rms deviation of about 2\%.

\begin{figure}[t]
\centerline{\epsfxsize=3.6in \epsffile{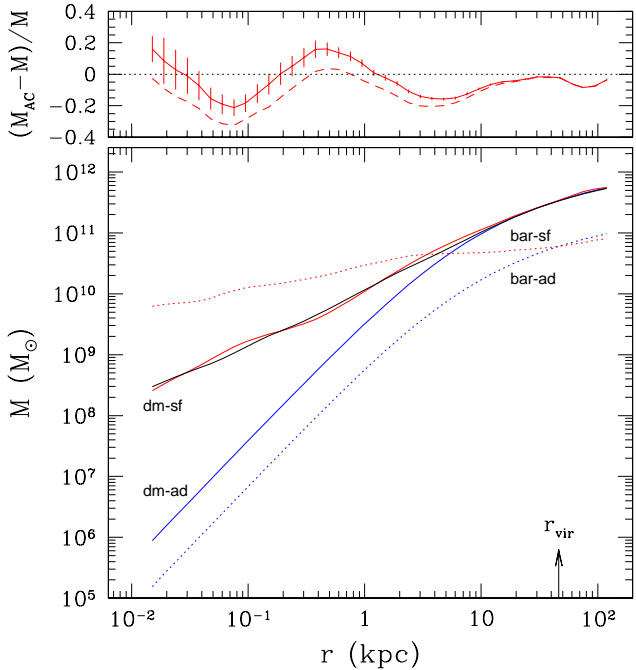}}
\vspace{-0.2cm}
\caption{Contraction of the dark matter profile in a simulated galaxy
  at $z=3$, from \citet{levine_etal08}. Line notation is the same as
  in Figure~\ref{fig:daisuke}.  Black solid line is the best-fit MAC
  model with $A_0=1.32$, $w=1.26$.  Vertical bars on the MAC model in
  the top panel indicate the Poisson uncertainty of the mass profile
  derived in the simulation.  Dashed line in top panel shows the MAC
  model prediction with fixed $A_0=1.6$, and best-fitting $w=1.26$.}
  \label{fig:robyn}
\vspace{0.2cm}
\end{figure}

\subsection{Galaxy Center}

Finally, we consider the resimulation of the galaxy run reported in
\citet{gnedin_etal04} that zooms into the innermost region of the
galaxy at $z=3$ \citep{levine_etal08}.  This simulation follows the
early evolution of a galaxy that becomes a Milky Way-sized object at
$z=0$.  The DM particle mass is $1.3\times 10^6\, \Msun$ and the peak
force resolution at $z=3$ is 0.064 kpc for the gas and 0.1 kpc for the
dark matter, a very small scale for cosmological simulations.  We
truncate the inner profile such that the innermost bin contains at
least 200 dark matter particles.

Figure~\ref{fig:robyn} shows that the MAC model is able to describe
even this case, with the rms deviation of 10\%.  This case is extreme
because the baryons dominate the dark matter by two orders of
magnitude at the innermost radius, and the dark matter mass is
enhanced by a factor of 300 relative to the extrapolation of the
dissipationless profile.

We also note that the stellar profile is contracted similarly to the
dark mater profile, because gas accretion is faster than star
formation.

\section{Distribution of model parameters}

\begin{figure}[t]
\vspace{-0.6cm}
\centerline{\epsfxsize=3.6in \epsffile{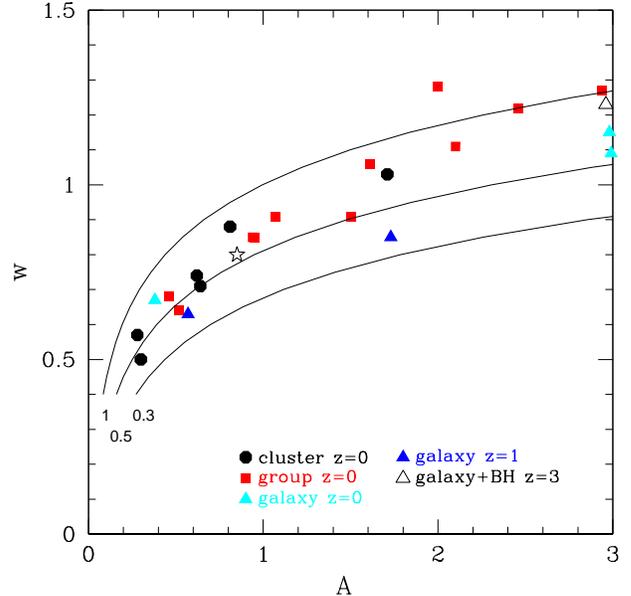}}
\vspace{-0.5cm}
\caption{Best-fitting parameters of the original MAC model
  (Equation~\ref{eq:rrave}) for all simulations discussed in this
  paper.  Asterisk marks the fiducial parameters of the MAC model in
  \citet{gnedin_etal04}.  Solid lines show the relation between $A$
  and $w$ that gives the same amount of contraction (enhancement of
  dark matter mass) at $r = 0.005 \, r_{\rm vir}$, for the baryon
  profile with $\nu=2$ normalized to equal the initial dark matter
  mass at $r_e = 0.05 \, r_{\rm vir}$.  The top line gives the same
  amount of contraction as the SAC model.  The other two lines
  correspond to 50\% and 30\% of that amount.}
  \label{fig:aw}
\vspace{0.2cm}
\end{figure}

All of the simulations considered here indicate some degree of
enhancement of the dark matter profile.  Not a single case indicates
halo expansion rather than contraction.  Figure~\ref{fig:aw}
combines the resulting constraints on the parameters $A$ and $w$ of
Equation (\ref{eq:rrave}).  The models do not fill all the available
parameter space, but instead concentrate in a fairly narrow region in
which $A$ and $w$ are strongly correlated.  The original MAC model
suggested by \citet{gnedin_etal04} falls right in the middle of the
new distribution.

\begin{figure}[t]
\vspace{-0.6cm}
\centerline{\epsfxsize=3.6in \epsffile{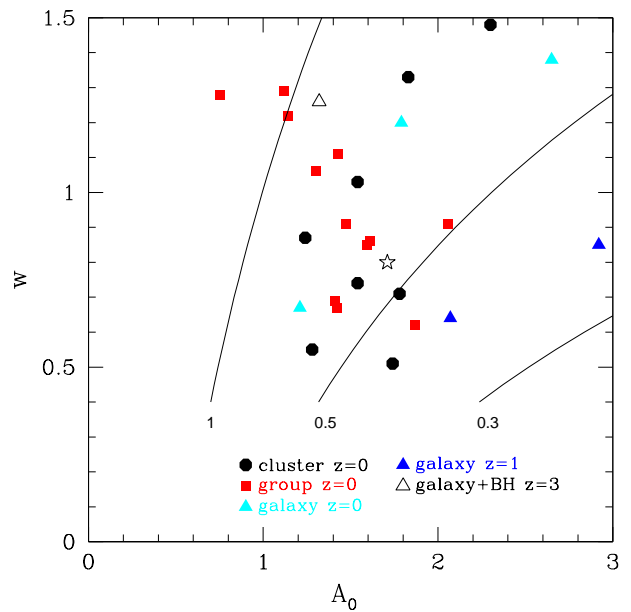}}
\vspace{-0.5cm}
\caption{Best-fitting parameters of the revised MAC model with $r_0 = 0.03 \,
  r_{\rm vir}$ (Equation~\ref{eq:rrave0}).  Symbols and lines are as
  in Figure~\ref{fig:aw}.}
  \label{fig:aw2}
\vspace{0.2cm}
\end{figure}

It is interesting to determine which combination of the parameters $A$
and $w$ yields the same amount of contraction.  Given the radial
dependence of the mass enhancement factor $F_M$
(Equation~\ref{eq:fm}), the solution to this problem varies with
radius.  However, we can remove most of the radial dependence by
defining the enhancement factor relative to the SAC model:
\begin{equation}
  f_M \equiv {F_M(r | A,w) \over F_M(r | 1,1)}
\end{equation}
and evaluating it at some inner radius where the linear approximation
for the contraction factor $y(r)$ is valid.  We take $r = 0.005 \, r_{\rm
vir}$, which corresponds to about 1~kpc for the Milky Way galaxy.  The
exact value of $r$ affects the resulting value of parameter $w$ (for a
given $A$) only logarithmically, as long as $r \ll r_e$, which we take
again to be $r_e = 0.05 \, r_{\rm vir}$.

Lines in Figure~\ref{fig:aw} show the relation between $A$ and $w$
corresponding to three values of $f_M$.  All simulations but three fall
below the level of contraction predicted by the SAC model.  At the
same time, no simulation falls below the level of $f_M = 0.3$.
Therefore, the MAC model is well constrained to be able to reliably
predict the amount of dark matter in the inner regions of galaxies and
clusters.

The isocontours of constant $f_M$ become even more horizontal at
smaller $r$.  This suggests that parameter $w$ may be more important
than $A$ in describing the amount of contraction.  It is also
desirable to describe the strength of the contraction effect by only
one parameter instead of two.  The first step in this direction is to
eliminate or reduce the apparent correlation between $A$ and $w$.  To
this aim, we calculated the best-fitting parameters of the revised
model (Equation~\ref{eq:rrave0}) and determined the pivot radius $r_0$
that minimizes the correlation.  It is the value $r_0 = 0.03 \, r_{\rm
vir}$ quoted above.  Figure~\ref{fig:aw2} shows the new distribution
of the resulting best-fit parameters $A_0$ and $w$.  The values of $w$
are essentially unchanged, but the values of $A_0$ are more
concentrated than the distribution of $A$ in Figure~\ref{fig:aw}.  The
residual scatter of $A_0$ reflects intrinsic variation of the strength
of halo contraction among different systems.  The isocontours of
constant $f_M$ also correspondingly change shape.

\begin{figure}[t]
\vspace{-0.6cm}
\centerline{\epsfxsize=3.6in \epsffile{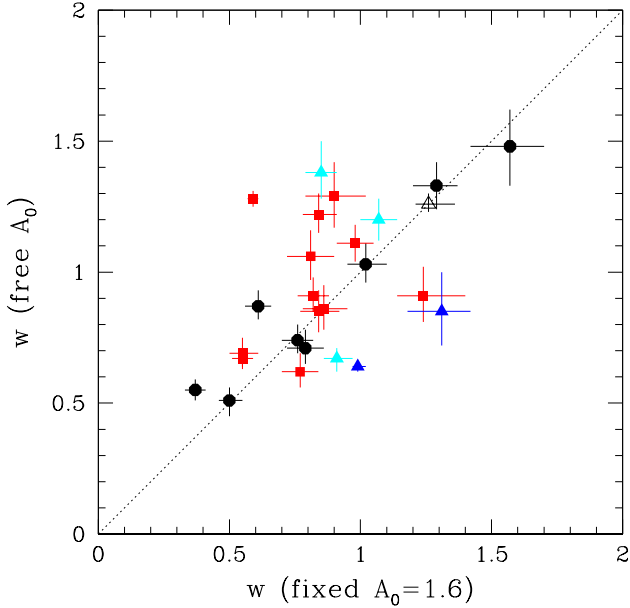}}
\vspace{-0.5cm}
\caption{Comparison of the best-fitting parameters $w$ obtained with
  variable $A_0$ with those obtained for fixed $A_0=1.6$.  Sizes of the
  errorbars are relative.}
  \label{fig:ww}
\vspace{0.2cm}
\end{figure}

The second step in simplifying the model prescription is fixing the
parameter $A_0$.  We take the average value $A_0 \approx 1.6$ and redo
all model fits allowing only for the variation of $w$.  The best-fit
values are listed in Table~\ref{tab:sim}.  Obviously, a one-parameter
fit is less accurate than the two-parameter fit, but the rms error of
mass is still typically below 10\%.  The one-parameter fits also do
not introduce systematic shifts in the derived values of $w$, as
shown in Figure~\ref{fig:ww}.  Thus, $w$ can serve as a convenient
measure of the strength of halo contraction.

\begin{figure}[t]
\vspace{-0.6cm}
\centerline{\epsfxsize=3.6in \epsffile{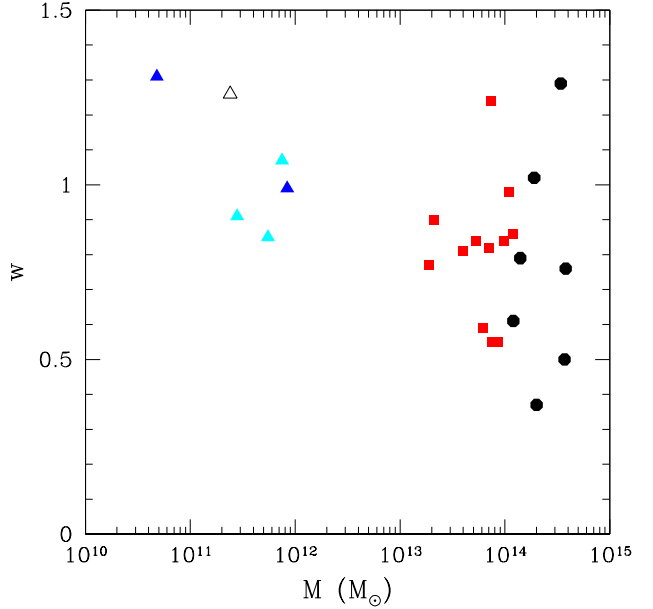}}
\vspace{-0.5cm}
\caption{Best-fitting parameter $w$ (for fixed $A_0 = 1.6$) vs. halo
  mass for all systems in our compilation.}
  \label{fig:w_m}
\vspace{0.2cm}
\end{figure}

\begin{figure}[t]
\vspace{-0.6cm}
\centerline{\epsfxsize=3.6in \epsffile{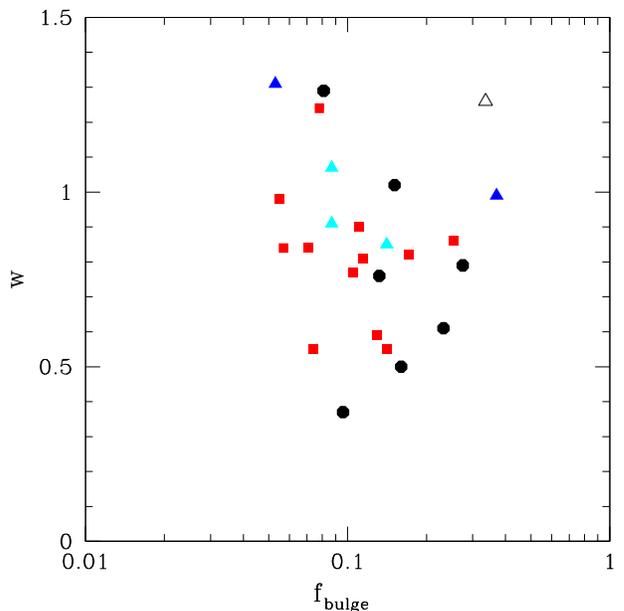}}
\vspace{-0.5cm}
\caption{Best-fitting parameter $w$ (for fixed $A_0 = 1.6$) vs. the
  bulge fraction, defined as the fraction of baryon mass contained
  within $0.01\, r_{\rm vir}$.}
  \label{fig:w_bul}
\vspace{0.2cm}
\end{figure}

In practical application of the contraction model to observations or
dissipationless simulations we wish to know the value of $w$ most
appropriate to a given system.  We considered several properties of
the simulated halos but, unfortunately, we were unable to find
significant correlation with $w$.  For example, Figure~\ref{fig:w_m}
shows that $w$ is effectively independent of halo mass.  Only the
lower envelope of the distribution decreases with $M_{\rm vir}$.

We have looked for other potential correlations: with the ratio of
final baryon mass to initial dark matter mass at the innermost
resolved radius and at a fixed radius of $0.01\, r_{\rm vir}$, with
the ratio of final dark matter to baryon mass at $0.01\, r_{\rm vir}$,
and with the bulge fraction of galaxies.  The latter is defined as the
fraction of baryon mass contained within $0.01\, r_{\rm vir}$:
$f_{\mathrm{bulge}} \equiv M_*(0.01\, r_{\rm vir})/M_*$.
Figure~\ref{fig:w_bul} shows the scatter plot of $w$ with the bulge
fraction, lacking any significant correlation.

\begin{figure}[t]
\vspace{-0.6cm}
\centerline{\epsfxsize=3.5in \epsffile{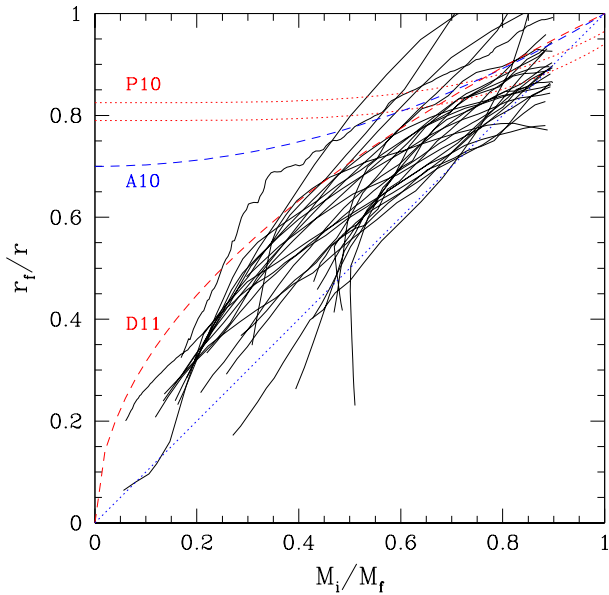}}
\vspace{-0.5cm}
\caption{Model-independent strength of the contraction effect.  Solid
  lines are for all systems in our compilation.  Dashed blue line is
  the relation suggested by \citet{abadi_etal10}, dotted lines are
  the relations suggested by \citet{pedrosa_etal10}, dashed red line
  is a relation of the type suggested by \citet{dutton_etal11} but
  with $n=0.5$ instead of $n=-0.5$ (which would go off the scale
  of the plot).}
  \label{fig:rfr}
\vspace{0.2cm}
\end{figure}

\section{Is halo contraction real?}
  \label{sec:discussion}

The strength of the contraction effect can be assessed in a
model-independent way, as suggested by \citet{abadi_etal10}.  They
expressed the ratio of the contracted to initial radius of spherical
shells as a function of the ratio of initial to final total mass.  In
the SAC model this relation is expected to be unity, $r_f/r =
M_i(r)/M_f(r_f)$.  For a weaker contraction effect, we expect $r_f/r >
M_i/M_f$.  \citet{abadi_etal10} found that in their simulations
the dark matter shells contract on the average as $r_f/r = 1 + 0.3
[(M_i/M_f)^2-1]$.  \citet{pedrosa_etal10} did a similar analysis for
their simulations but found different relations: $r_f/r = 1 + 0.14
[(M_i/M_f)^4-1.25]$ for early-type galaxies, and $r_f/r = 1 + 0.15
[(M_i/M_f)^4-1.4]$ for late-type galaxies.

Figure~\ref{fig:rfr} shows the contraction factor for all of our
simulations.  The effect is clearly more significant than suggested by
the above relations.  In several cases, the contraction is even
stronger at the innermost radii than that predicted by the SAC model,
that is $r_f/r < M_i/M_f$.  It can also be seen in
Figure~\ref{fig:aw2}.  Overall, there is no single relation for the
contraction factor $r_f/r$.  The scatter among different systems is
intrinsic, similar to the scatter of model parameters $A$ and $w$.

\citet{dutton_etal11} constructed semi-analytical models of galaxy
populations using, in addition to the MAC model, an analytic
prescription for either halo contraction or expansion of the form
$r_f/r = (M_i/M_f)^n$, with $n$ as a free parameter.  Their preferred
case is halo expansion with $n=-0.5$.  None of our systems shows
evidence for such expansion.  If we were to use this parametrization
to envelop the distribution of our simulations, then most of our
results would lie within the range $0.6 < n < 1.2$.  However, such
parametrization has no physical basis and we do not recommend it.

Is the halo contraction effect real?  Much of the apparent controversy
in the literature on the validity of the contraction effect is due to
the strict application of adiabatic invariants.  In fact, most of
the hydrodynamic cosmological simulations are in agreement that the
effect is real and, at the same time, weaker than that expected in the
\citet{blumenthal_etal86} model.  To avoid future controversy, we
propose to abandon the term ``adiabatic contraction'' (reserving it
only for the SAC model for historical reasons) and instead use the
term ``halo contraction''.

Note that all individual effects that were proposed to ``reverse''
contraction (such as the rapid supernova winds, cold accretion,
galactic bars, inspiraling of dense baryonic clumps by dynamical
friction, etc.) are already included in self-consistent cosmological
simulations and differ only in the specific implementation (numerical
resolution, star formation prescription, feedback model).  Isolated
investigations of these effects therefore do not invalidate the
conclusions we derive from the ensemble of the simulations presented
here.  One should not also attribute halo contraction only to
dissipative galaxy formation and contrast it with dissipationless
accretion of satellites \citep[e.g.,][]{lackner_ostriker10} because
both processes are taking place simultaneously.  Hydrodynamic
simulations are steadily improving in accuracy and will continue to
include and evaluate new effects, such as the repeated potential
fluctuations following bursts of star formation
\citep{pontzen_governato11}, mechanical and radiative feedback of
active galactic nuclei, etc.  Our ability to model halo contraction
will continue to evolve along with our overall understanding of galaxy
formation.

The MAC model does not presume a specific amount of contraction.  It
is a method for calculating the dynamical response of the dark matter
halo to a given accumulated mass of baryons.  The model is based on
simple underlying physics and not just fitting the results of a
particular simulation.  The halo response depends directly on the
change of the baryon profile relative to the dissipationless
formation.  As such, the response may be a halo expansion if more
baryons are removed from the galaxy center than were there initially.
For example, \citet{sommer-larsen_limousin10} gradually reduced the
stellar mass in their simulated galaxy cluster between $z=2$ and
$z=0$, and found a much reduced contraction effect at the end, as
would be expected if the removed stars did not form at all.

How should halo contraction be included in modeling of observed
systems or in theoretical semi-analytical models based on
dissipationless simulations?  Direct parametrization of the form
suggested by \citet{abadi_etal10} can only be used when the final
distribution of dark matter is known, that is, only for comparing
results of hydrodynamic simulations with each other.  When the
distribution of dark matter needs to be predicted, the MAC model
provides a reasonably accurate method and requires knowing only the
final baryon profile.  Unfortunately, the values of the model
parameters have irreducible scatter and do not appear to correlate
with any obvious property of the system of interest.  Thus the model
prediction carries some irreducible uncertainty.  In practical
application it suffices to account for this uncertainty by varying
only one of the two parameters.  We advocate the choice of $A_0$ and
$w$ (Equation~\ref{eq:rrave0}) instead of $A$ and $w$
(Equation~\ref{eq:rrave}) because the new parameters are effectively
uncorrelated.  For assessing the effect of halo contraction on a given
system, we suggest fixing $A_0 = 1.6$ and varying $w$ in the range
$0.6 - 1.3$.  This prescription covers most of the parameter range
seen in our simulations (Figure~\ref{fig:aw2}) while providing the
estimate of the dark matter mass profile with rms accuracy of about
10\% (Table~\ref{tab:sim}).

What is the origin of the intrinsic scatter of the parameters $A_0$
and $w$, or of the contraction factor $r_f/r$?  We can identify a
number of processes in galaxy formation that could produce such scatter.
For example, baryons in different galaxies condense at different
epochs and have different angular momentum profiles, both of which
affect the compactness of the final baryon distribution.  In addition,
galaxies experience different amounts of merger activity, which can
change the morphology of baryon distribution and can gravitationally
heat dark matter.  Given these and other factors, it would be
surprising if the intrinsic scatter did not exist.  It is not known at
the moment which of these effects are dominant.  It would be useful to
explore the origin of the intrinsic scatter in future studies.

\section{Conclusions}

We have evaluated the halo contraction effect using a large
compilation of cosmological hydrodynamic simulations performed by
different groups with different codes.  In all the cases we
considered, we find an increase of the dark matter density in inner
regions of galaxies and clusters, relative to the matching
dissipationless simulation.  The halo contraction effect is real and
must be included in the modeling of observations and in the
semi-analytical theoretical modeling.

The contraction effect is weaker than predicted by the adiabatic
contraction model of \citet{blumenthal_etal86}.  However, depending on
the system and the final baryon distribution, the inner dark matter
density is still enhanced typically by a factor of several, and in
extreme cases by two orders of magnitude.

The revised MAC model offers a convenient and accurate way to estimate
the effect of halo contraction.  The distribution of model parameters
cannot be reduced to a single number, but their range is well
constrained: $1 < A_0 < 2.2$, $0.6 < w < 1.3$.  We find that fixing
the value of $A_0 = 1.6$ does not significantly degrade the accuracy
of the predicted mass profile, relative to the two-parameter fit.  We
suggest varying $w$ in the range $0.6 - 1.3$ in order to bracket a
possible response of the dark matter halo in a given system of
interest.

The revised MAC model is encoded in the software package {\it Contra},
available for download at
http://www.astro.lsa.umich.edu/$\sim$ognedin/contra.

\acknowledgements 

We thank Mario Abadi and Alan Duffy for clarifying discussions and
sending us electronic data from their numerical simulations, and Fabio
Governato and Andrew Pontzen for additional discussions.  O.Y.G.,
N.Y.G, and A.V.K. were supported in part by the National Science
Foundation through grant AST-0708087.  O.Y.G. was also supported in
part by NASA through HST programs GO-10824 and GO-11589.  D.N. was
supported in part by the NSF grant AST-1009811, by NASA ATP grant
NNX11AE07G, and by the Yale University.  G.Y. acknowledges financial
support from MICINN (Spain) through research grants FPA2009-08958,
AYA2009-13875-C03-02, and Consolider-Ingenio SyeC (CSD2007-0050).
D.C. is a Golda-Meir fellow.

\appendix
\section{Analytical approximation for the contraction model}

An approximate analytical solution of Equation~(\ref{eq:yi}) for the
contraction factor $y \equiv r_f/r$ can be obtained as follows.  At $r
\ll r_e$, the ratio $M_b(\bar{r})/M_i(\bar{r}) \gg 1$ and the term
$(1-f_b)$ can be neglected.  The first-order approximation is
therefore
\begin{equation}
  y_0 = \left[{ M_b(\bar{r}) \over M_i(\bar{r}) }\right]^{-{1\over 1+w(3-\nu)}}.
  \label{eq:y0}
\end{equation}
Then we can express the correct solution as $y = y_0 (1+\delta)$, with
$\delta \la 1$.  Substituting this into Equation~(\ref{eq:yi}) and
expanding the power term of $(1+\delta)$, we obtain
\begin{equation}
  \delta = {(1-f_b) y_0 \over 1+w(3-\nu) + (1-f_b) y_0}.
  \label{eq:y1} 
\end{equation}
The value of $\delta$ is of the order $y_0$, which means that this approximation
is valid where $y_0 \la 1$, or $r \la r_e$.

To evaluate the mass enhancement factor $F_M(r)$ at a specified radius
$r$, we must express the initial radius $r$ as a function of the
contracted radius $ry(r)$.  Considering only the first-order
approximation $y \approx y_0$, we find a power-law solution $ry
\propto r^\alpha$, where $\alpha = (1+2w)/(1+w(3-\nu))$.  Then
$F_M(ry) \propto (ry)^{\frac{2}{\alpha}-2}$.  Retaining all the
coefficients, we have
\begin{equation}
  F_M(ry) \approx \left[{ M_b(\bar{r}) \over M_i(\bar{r}) }\right]^{2 \over 1+2w} = 
    \left[{ (1-f_b) \, \left({\bar{r} \over r_e}\right)^{1-\nu} }\right]^{2 \over 1+2w}.
  \label{eq:fma}
\end{equation}
This approximation is valid at $r \ll r_e$.

Using the same first-order approximation, we can derive the inner
logarithmic slope of the contracted dark matter profile, $\rho_{\rm
dm}(r) \propto r^{-\gamma}$:
\begin{equation}
  \gamma = {1 + 2 w \nu \over 1 + 2 w}.
\end{equation}
This slope was already derived as Equation~(A12) in
\citet{gnedin_etal04}.  This equation links the contracted dark matter
slope with the slope of the baryon profile.  It provides an accurate
description of the dark matter profile in the hydrodynamic simulations
described above, within the errors of calculation of $w$ and $\nu$.

\bibliography{gc,contra}

\begin{thebibliography}{53}
\expandafter\ifx\csname natexlab\endcsname\relax\def\natexlab#1{#1}\fi

\bibitem[{{Abadi} {et~al.}(2010){Abadi}, {Navarro}, {Fardal}, {Babul}, \&
  {Steinmetz}}]{abadi_etal10}
{Abadi}, M.~G., {Navarro}, J.~F., {Fardal}, M., {Babul}, A., \& {Steinmetz}, M.
  2010, \mnras, 407, 435

\bibitem[{{Abadi} {et~al.}(2003){Abadi}, {Navarro}, {Steinmetz}, \&
  {Eke}}]{abadi_etal03}
{Abadi}, M.~G., {Navarro}, J.~F., {Steinmetz}, M., \& {Eke}, V.~R. 2003, \apj,
  591, 499

\bibitem[{{Auger} {et~al.}(2010){Auger}, {Treu}, {Gavazzi}, {Bolton},
  {Koopmans}, \& {Marshall}}]{auger_etal10}
{Auger}, M.~W., {Treu}, T., {Gavazzi}, R., {Bolton}, A.~S., {Koopmans},
  L.~V.~E., \& {Marshall}, P.~J. 2010, \apjl, 721, L163

\bibitem[{{Barnes} \& {White}(1984)}]{barnes_white84}
{Barnes}, J. \& {White}, S.~D.~M. 1984, \mnras, 211, 753

\bibitem[{{Benson} \& {Bower}(2010)}]{benson_bower10}
{Benson}, A.~J. \& {Bower}, R. 2010, \mnras, 405, 1573

\bibitem[{{Blumenthal} {et~al.}(1986){Blumenthal}, {Faber}, {Flores}, \&
  {Primack}}]{blumenthal_etal86}
{Blumenthal}, G.~R., {Faber}, S.~M., {Flores}, R., \& {Primack}, J.~R. 1986,
  \apj, 301, 27

\bibitem[{{Ceverino} \& {Klypin}(2009)}]{ceverino_klypin09}
{Ceverino}, D. \& {Klypin}, A. 2009, \apj, 695, 292

\bibitem[{{Choi} {et~al.}(2006){Choi}, {Lu}, {Mo}, \& {Weinberg}}]{choi_etal06}
{Choi}, J., {Lu}, Y., {Mo}, H.~J., \& {Weinberg}, M.~D. 2006, \mnras, 372, 1869

\bibitem[{{Col{\'{\i}}n} {et~al.}(2006){Col{\'{\i}}n}, {Valenzuela}, \&
  {Klypin}}]{colin_etal06}
{Col{\'{\i}}n}, P., {Valenzuela}, O., \& {Klypin}, A. 2006, \apj, 644, 687

\bibitem[{{Debattista} {et~al.}(2008){Debattista}, {Moore}, {Quinn},
  {Kazantzidis}, {Maas}, {Mayer}, {Read}, \& {Stadel}}]{debattista_etal08}
{Debattista}, V.~P., {Moore}, B., {Quinn}, T., {Kazantzidis}, S., {Maas}, R.,
  {Mayer}, L., {Read}, J., \& {Stadel}, J. 2008, \apj, 681, 1076

\bibitem[{{D{\'e}mocl{\`e}s} {et~al.}(2010){D{\'e}mocl{\`e}s}, {Pratt},
  {Pierini}, {Arnaud}, {Zibetti}, \& {D'Onghia}}]{democles_etal10}
{D{\'e}mocl{\`e}s}, J., {Pratt}, G.~W., {Pierini}, D., {Arnaud}, M., {Zibetti},
  S., \& {D'Onghia}, E. 2010, \aap, 517, A52

\bibitem[{{Dubinski} \& {Carlberg}(1991)}]{dubinski_carlberg91}
{Dubinski}, J. \& {Carlberg}, R.~G. 1991, \apj, 378, 496

\bibitem[{{Duffy} {et~al.}(2010){Duffy}, {Schaye}, {Kay}, {Dalla Vecchia},
  {Battye}, \& {Booth}}]{duffy_etal10}
{Duffy}, A.~R., {Schaye}, J., {Kay}, S.~T., {Dalla Vecchia}, C., {Battye},
  R.~A., \& {Booth}, C.~M. 2010, \mnras, 405, 2161

\bibitem[{{Dutton} {et~al.}(2011){Dutton}, {Conroy}, {van den Bosch}, {Simard},
  {Mendel}, {Courteau}, {Dekel}, {More}, \& {Prada}}]{dutton_etal11}
{Dutton}, A.~A., {Conroy}, C., {van den Bosch}, F.~C., {Simard}, L., {Mendel},
  J.~T., {Courteau}, S., {Dekel}, A., {More}, S., \& {Prada}, F. 2011, \mnras,
  1045

\bibitem[{{Dutton} {et~al.}(2007){Dutton}, {van den Bosch}, {Dekel}, \&
  {Courteau}}]{dutton_etal07}
{Dutton}, A.~A., {van den Bosch}, F.~C., {Dekel}, A., \& {Courteau}, S. 2007,
  \apj, 654, 27

\bibitem[{{Eggen} {et~al.}(1962){Eggen}, {Lynden-Bell}, \&
  {Sandage}}]{eggen_etal62}
{Eggen}, O.~J., {Lynden-Bell}, D., \& {Sandage}, A.~R. 1962, \apj, 136, 748

\bibitem[{{Gnedin} {et~al.}(2004){Gnedin}, {Kravtsov}, {Klypin}, \&
  {Nagai}}]{gnedin_etal04}
{Gnedin}, O.~Y., {Kravtsov}, A.~V., {Klypin}, A.~A., \& {Nagai}, D. 2004, \apj,
  616, 16

\bibitem[{{Gnedin} \& {Zhao}(2002)}]{gnedin_zhao02}
{Gnedin}, O.~Y. \& {Zhao}, H. 2002, \mnras, 333, 299

\bibitem[{{Gottloeber} {et~al.}(2010){Gottloeber}, {Hoffman}, \&
  {Yepes}}]{gottlober_etal10}
{Gottloeber}, S., {Hoffman}, Y., \& {Yepes}, G. 2010, arXiv:1005.2687

\bibitem[{{Governato} {et~al.}(2010){Governato}, {Brook}, {Mayer}, {Brooks},
  {Rhee}, {Wadsley}, {Jonsson}, {Willman}, {Stinson}, {Quinn}, \&
  {Madau}}]{governato_etal10}
{Governato}, F., {Brook}, C., {Mayer}, L., {Brooks}, A., {Rhee}, G., {Wadsley},
  J., {Jonsson}, P., {Willman}, B., {Stinson}, G., {Quinn}, T., \& {Madau}, P.
  2010, \nat, 463, 203

\bibitem[{{Governato} {et~al.}(2007){Governato}, {Willman}, {Mayer}, {Brooks},
  {Stinson}, {Valenzuela}, {Wadsley}, \& {Quinn}}]{governato_etal07}
{Governato}, F., {Willman}, B., {Mayer}, L., {Brooks}, A., {Stinson}, G.,
  {Valenzuela}, O., {Wadsley}, J., \& {Quinn}, T. 2007, \mnras, 374, 1479

\bibitem[{{Guedes} {et~al.}(2011){Guedes}, {Callegari}, {Madau}, \&
  {Mayer}}]{guedes_etal11}
{Guedes}, J., {Callegari}, S., {Madau}, P., \& {Mayer}, L. 2011, \apj, in
  press; arXiv:1103.6030

\bibitem[{{Gustafsson} {et~al.}(2006){Gustafsson}, {Fairbairn}, \&
  {Sommer-Larsen}}]{gustafsson_etal06}
{Gustafsson}, M., {Fairbairn}, M., \& {Sommer-Larsen}, J. 2006, \prd, 74,
  123522

\bibitem[{{Johansson} {et~al.}(2009){Johansson}, {Naab}, \&
  {Ostriker}}]{johansson_etal09}
{Johansson}, P.~H., {Naab}, T., \& {Ostriker}, J.~P. 2009, \apjl, 697, L38

\bibitem[{{Kazantzidis} {et~al.}(2004){Kazantzidis}, {Kravtsov}, {Zentner},
  {Allgood}, {Nagai}, \& {Moore}}]{kazantzidis_etal04}
{Kazantzidis}, S., {Kravtsov}, A.~V., {Zentner}, A.~R., {Allgood}, B., {Nagai},
  D., \& {Moore}, B. 2004, \apjl, 611, L73

\bibitem[{{Knebe} {et~al.}(2010){Knebe}, {Libeskind}, {Knollmann}, {Yepes},
  {Gottl{\"o}ber}, \& {Hoffman}}]{knebe_etal10}
{Knebe}, A., {Libeskind}, N.~I., {Knollmann}, S.~R., {Yepes}, G.,
  {Gottl{\"o}ber}, S., \& {Hoffman}, Y. 2010, \mnras, 405, 1119

\bibitem[{{Koopmans} {et~al.}(2009){Koopmans}, {Bolton}, {Treu}, {Czoske},
  {Auger}, {Barnab{\`e}}, {Vegetti}, {Gavazzi}, {Moustakas}, \&
  {Burles}}]{koopmans_etal09}
{Koopmans}, L.~V.~E., {Bolton}, A., {Treu}, T., {Czoske}, O., {Auger}, M.~W.,
  {Barnab{\`e}}, M., {Vegetti}, S., {Gavazzi}, R., {Moustakas}, L.~A., \&
  {Burles}, S. 2009, \apjl, 703, L51

\bibitem[{{Kravtsov}(1999)}]{kravtsov99}
{Kravtsov}, A.~V. 1999, PhD thesis, New Mexico State University

\bibitem[{{Kravtsov} {et~al.}(2002){Kravtsov}, {Klypin}, \&
  {Hoffman}}]{kravtsov_etal02}
{Kravtsov}, A.~V., {Klypin}, A., \& {Hoffman}, Y. 2002, \apj, 571, 563

\bibitem[{{Lackner} \& {Ostriker}(2010)}]{lackner_ostriker10}
{Lackner}, C.~N. \& {Ostriker}, J.~P. 2010, \apj, 712, 88

\bibitem[{{Levine} {et~al.}(2008){Levine}, {Gnedin}, {Hamilton}, \&
  {Kravtsov}}]{levine_etal08}
{Levine}, R., {Gnedin}, N.~Y., {Hamilton}, A.~J.~S., \& {Kravtsov}, A.~V. 2008,
  \apj, 678, 154

\bibitem[{{Meza} {et~al.}(2003){Meza}, {Navarro}, {Steinmetz}, \&
  {Eke}}]{meza_etal03}
{Meza}, A., {Navarro}, J.~F., {Steinmetz}, M., \& {Eke}, V.~R. 2003, \apj, 590,
  619

\bibitem[{{Minor} \& {Kaplinghat}(2008)}]{minor_kaplinghat08}
{Minor}, Q.~E. \& {Kaplinghat}, M. 2008, \mnras, 391, 653

\bibitem[{{Moore} {et~al.}(1998){Moore}, {Governato}, {Quinn}, {Stadel}, \&
  {Lake}}]{moore_etal98}
{Moore}, B., {Governato}, F., {Quinn}, T., {Stadel}, J., \& {Lake}, G. 1998,
  \apjl, 499, L5

\bibitem[{{Nagai}(2006)}]{nagai06}
{Nagai}, D. 2006, \apj, 650, 538

\bibitem[{{Napolitano} {et~al.}(2011){Napolitano}, {Romanowsky}, {Capaccioli},
  {Douglas}, {Arnaboldi}, {Coccato}, {Gerhard}, {Kuijken}, {Merrifield},
  {Bamford}, {Cortesi}, {Das}, \& {Freeman}}]{napolitano_etal11}
{Napolitano}, N.~R., {Romanowsky}, A.~J., {Capaccioli}, M., {Douglas}, N.~G.,
  {Arnaboldi}, M., {Coccato}, L., {Gerhard}, O., {Kuijken}, K., {Merrifield},
  M.~R., {Bamford}, S.~P., {Cortesi}, A., {Das}, P., \& {Freeman}, K.~C. 2011,
  \mnras, 411, 2035

\bibitem[{{Navarro} {et~al.}(1997){Navarro}, {Frenk}, \&
  {White}}]{navarro_etal97}
{Navarro}, J.~F., {Frenk}, C.~S., \& {White}, S.~D.~M. 1997, \apj, 490, 493

\bibitem[{{Navarro} {et~al.}(2010){Navarro}, {Ludlow}, {Springel}, {Wang},
  {Vogelsberger}, {White}, {Jenkins}, {Frenk}, \& {Helmi}}]{navarro_etal10}
{Navarro}, J.~F., {Ludlow}, A., {Springel}, V., {Wang}, J., {Vogelsberger}, M.,
  {White}, S.~D.~M., {Jenkins}, A., {Frenk}, C.~S., \& {Helmi}, A. 2010,
  \mnras, 402, 21

\bibitem[{{Pedrosa} {et~al.}(2009){Pedrosa}, {Tissera}, \&
  {Scannapieco}}]{pedrosa_etal09}
{Pedrosa}, S., {Tissera}, P.~B., \& {Scannapieco}, C. 2009, \mnras, 395, L57

\bibitem[{{Pedrosa} {et~al.}(2010){Pedrosa}, {Tissera}, \&
  {Scannapieco}}]{pedrosa_etal10}
---. 2010, \mnras, 402, 776

\bibitem[{{Pontzen} \& {Governato}(2011)}]{pontzen_governato11}
{Pontzen}, A. \& {Governato}, F. 2011, \apj, submitted; arXiv:1106.0499

\bibitem[{{Romano-D{\'{\i}}az} {et~al.}(2008){Romano-D{\'{\i}}az}, {Shlosman},
  {Hoffman}, \& {Heller}}]{romanodiaz_etal08}
{Romano-D{\'{\i}}az}, E., {Shlosman}, I., {Hoffman}, Y., \& {Heller}, C. 2008,
  \apjl, 685, L105

\bibitem[{{Rozo} {et~al.}(2008){Rozo}, {Nagai}, {Keeton}, \&
  {Kravtsov}}]{rozo_etal08}
{Rozo}, E., {Nagai}, D., {Keeton}, C., \& {Kravtsov}, A. 2008, \apj, 687, 22

\bibitem[{{Ryden} \& {Gunn}(1987)}]{ryden_gunn87}
{Ryden}, B.~S. \& {Gunn}, J.~E. 1987, \apj, 318, 15

\bibitem[{{Schulz} {et~al.}(2010){Schulz}, {Mandelbaum}, \&
  {Padmanabhan}}]{schulz_etal10}
{Schulz}, A.~E., {Mandelbaum}, R., \& {Padmanabhan}, N. 2010, \mnras, 408, 1463

\bibitem[{{Seigar} {et~al.}(2008){Seigar}, {Barth}, \&
  {Bullock}}]{seigar_etal08}
{Seigar}, M.~S., {Barth}, A.~J., \& {Bullock}, J.~S. 2008, \mnras, 389, 1911

\bibitem[{{Sellwood} \& {McGaugh}(2005)}]{sellwood_mcgaugh05}
{Sellwood}, J.~A. \& {McGaugh}, S.~S. 2005, \apj, 634, 70

\bibitem[{{Sommer-Larsen} \& {Limousin}(2010)}]{sommer-larsen_limousin10}
{Sommer-Larsen}, J. \& {Limousin}, M. 2010, \mnras, 408, 1998

\bibitem[{{Tissera} {et~al.}(2010){Tissera}, {White}, {Pedrosa}, \&
  {Scannapieco}}]{tissera_etal10}
{Tissera}, P.~B., {White}, S.~D.~M., {Pedrosa}, S., \& {Scannapieco}, C. 2010,
  \mnras, 406, 922

\bibitem[{{Vikhlinin} {et~al.}(1999){Vikhlinin}, {McNamara}, {Hornstrup},
  {Quintana}, {Forman}, {Jones}, \& {Way}}]{vikhlinin_etal99}
{Vikhlinin}, A., {McNamara}, B.~R., {Hornstrup}, A., {Quintana}, H., {Forman},
  W., {Jones}, C., \& {Way}, M. 1999, \apjl, 520, L1

\bibitem[{{Zappacosta} {et~al.}(2006){Zappacosta}, {Buote}, {Gastaldello},
  {Humphrey}, {Bullock}, {Brighenti}, \& {Mathews}}]{zappacosta_etal06}
{Zappacosta}, L., {Buote}, D.~A., {Gastaldello}, F., {Humphrey}, P.~J.,
  {Bullock}, J., {Brighenti}, F., \& {Mathews}, W. 2006, \apj, 650, 777

\bibitem[{{Zeldovich} {et~al.}(1980){Zeldovich}, {Klypin}, {Khlopov}, \&
  {Chechetkin}}]{zeldovich_etal80}
{Zeldovich}, Y.~B., {Klypin}, A.~A., {Khlopov}, M.~Y., \& {Chechetkin}, V.~M.
  1980, Soviet J. Nucl. Phys., 31, 664

\bibitem[{{Zemp} {et~al.}(2011){Zemp}, {Gnedin}, {Gnedin}, \&
  {Kravtsov}}]{zemp_etal11}
{Zemp}, M., {Gnedin}, O.~Y., {Gnedin}, N.~Y., \& {Kravtsov}, A.~V. 2011, \apj,
  submitted

\end{thebibliography}

\end{document}